\documentclass[12pt, draftclsnofoot, onecolumn]{IEEEtran}
\usepackage{cite}

\ifCLASSINFOpdf
  \usepackage[pdftex]{graphicx}
\else
  \usepackage[dvips]{graphicx}
\fi
\usepackage[cmex10]{amsmath}
\usepackage{amssymb}
\usepackage{amsthm}
\usepackage{color}

\usepackage{array}

\usepackage{booktabs}
\usepackage{multirow}
\usepackage{rotating}

\usepackage{color}

\begin{document}
%
\title{On-Board Beam Generation for Multibeam  Satellite Systems}
%
%
%

\author{Vahid Joroughi,
        Miguel \'Angel V\'azquez
        Ana I. P\'erez-Neira,~\IEEEmembership{Senior Member,~IEEE} and Bertrand~Devillers
        \thanks{This work has been partially presented in IEEE Globecom 2013}
        \thanks{This work was performed in the context of the framework of SatNEx III
through ESA. The research leading to these results has received funding from the Spanish Ministry of Science and Innovation under projects TEC2011-29006-C03-02 (GRE3N-LINK-MAC) and the Catalan Government (2014 SGR 1567).}
\thanks{V. Joroughi,  M. \'A. V\'azquez and A. P\'erez-Neira are with the Centre Tecnol\`ogic de les Telecomunicacions de 
Catalunya (CTTC), Barcelona, Spain. A. P\'erez-Neira and V. Joroughi are also with  Universitat Politecnica de Catalunya (UPC).}
\thanks{Emails:vahid.joroughi@upc.edu,~mavazquez@cttc.es,~ana.isabel.perez@upc.edu}
\thanks{B. Devillers is with European Patent Office (EPO), Netherlands}
\thanks{Email:bertrand.devillers@gmail.com}}

%
%

\markboth{Submitted}%
{}
%



\maketitle

\begin{abstract}
This paper aims at designing an on-board beam generation process for multibeam satellite systems with the goal of reducing the traffic at the feeder link. Full frequency reuse among beams is considered and the beamforming at the satellite is designed for supporting interference mitigation techniques. In addition, in order to reduce the payload cost and complexity, this on-board processing is assumed to be constant and the same for forward and return link transmissions. To meet all these requirements a novel robust minimum mean square error (MMSE) optimization is conceived. The benefits of the considered scheme are evaluated with respect to the current approaches both analytically and numerically. Indeed, we show that with the DVB-RCS and DVB-S2 standards, our proposal allows to increase the total throughput within a range between 6\% and 15\% with respect to other on-board processing techniques in the return and forward link, respectively. Furthermore, the proposed solution presents an implicit feeder link bandwidth reduction with respect to the on-ground beam generation process.
\end{abstract}


%
\IEEEpeerreviewmaketitle

\section{Introduction}
%
%
%
%
\IEEEPARstart{T}{he} increasing demand for fixed broadband data services is an opportunity for satellite industries to target new markets apart from the well-known current ones (i.e. broadcast broadband, emergency communications, ...). In order to cope with higher data traffic demands, satellite system designers are looking for advanced satellite communication architectures. In this context, the use of multiple beams  has recently received a lot of attention as a key enabler of next generation high throughput satellite systems. These systems rely on implementing a large number of beams instead of a single (global) beam in  the coverage area. This is beneficial {\color{black} since each beam can have a larger antenna gain-to-noise temperature than in the single beam case and the available spectrum can be reused among spatially separated beams. Furthermore, whenever the satellite systems delivers broadband interactive traffic, the multibeam architecture can support different modulations and code rates to each beam depending on the user link quality, leading to a high increase of the overall system throughput.}

The implementation of multibeam satellite systems is currently being investigated in order to increase the overall spectral efficiency while keeping the payload complexity affordable. One of the main challenges is how to deal with the large spectral demands of the feeder link (i.e. the bidirectional communication link between the satellite and the service provider), whose bandwidth requirements  increase exponentially as it aggregates the traffic of all users. Recently, some techniques have appeared in order to reduce the feeder link spectrum requirements. There is a current tendency for moving the feeder from the Ka band to the Q/V band, where there are larger available bandwidths \cite{Gayrard09}. Unfortunately, in these frequencies the fading is extremely large and more advanced transmitting diversity techniques are needed. 

Another option is the use of multiple gateways, which might be adequate in order to reduce the feeder link spectral requirements  as they can be equipped with very directive antennas and exploit the spatial diversity while sharing all available spectrum \cite{Zheng2012,Gharanjik2013}. Nevertheless, the deployment of several gateways increases the cost of the system and; moreover, the interference mitigation techniques suffer from certain degradation \cite{vahid14,zheng12}. This is due to the fact that the processing must be separated in isolated processing units.

In contrast to the aforementioned feeder link traffic reduction techniques, this paper focuses on the on-board beam generation process. This promising solution keeps certain processing in the payload so that the amount of required signals from the feeder link are severally reduced. In this way, the satellite does not act in transparent mode and it carries out some processing, leading to a high reduction of the feeder link bandwidth requirements. Specifically, while the on-ground beamforming requires a feeder link bandwidth of
\begin{equation}
B_{\text{feeder link on-ground}} = NB_{\text{user}},
\end{equation}
where $N$ is the number of feed elements\footnote{The input signals of the antenna array feed assembly located in the payload.} and $B_{\text{user}}$ is the user bandwidth; the on-board beamforming only requires
\begin{equation}
B_{\text{feeder link on-board}} = KB_{\text{user}},
\end{equation}
where $K$ is the number of users. {\color{black} For this work we will consider multiple-feed-per-beam architecture where  $N > K$. Note that, in contrast to single-feed-per-beam architectures ($N = K$), in multiple-feed-per-beam architectures beamforming scan losses are negligible \cite{mfpb2011}.} A more detailed description of the beam process is presented in \cite{Tronc2013}.

Apart from the feeder link challenge, multibeam satellite systems require a large capacity in the access network. As a matter of fact, in the generated radiation pattern on Earth, adjacent beams create high levels of interference and, therefore, a carefully planned power and frequency reuse among beams must be employed to cope with this increased level of interference. Consequently, beams with adjacent footprint currently operate in different frequency bands or polarizations. In this context, the number of colors $N_c$ is the  essential parameter, which corresponds to the number of disjoint frequency bands and polarizations employed on the coverage area ($N_c\geq 1$). In fact, the lower the number of $N_c$, the higher the overall system bandwidth will be and the higher the interference power levels will be generated. 

In order to increase the available bandwidth yet maintaining a low multiuser interference, a promising technique is to use full frequency reuse pattern ($N_c=1$) and resort to interference mitigation techniques.  In this way, signals can be precoded and detected before being transmitted and received  in order to reduce inter-beam interference \cite{Gallinaro08}. As a result, a considerable improvement of the achievable spectral efficiency can be obtained. To this end, more advanced interference mitigation techniques as precoding in the forward link and multiuser detection or filtering in the return link have been considered in past studies of the European Space Agency (ESA)\cite{Gallinaro08},\cite{Boussemart2011}.

Since interference mitigation techniques require large computational resources, they must be carried out on ground. Indeed, larger efficiencies are obtained if not only the precoding and detection are done on ground, but also the beam generation process, as more flexible processing units are available. In other words, if the beamforming is kept fixed on the payload, there is a performance loss compared to the spectral efficiencies obtained by on ground beamforming \cite{bertrand,Devillers2011}. However, if the satellite does not perform any beam processing, the feeder link needs a large amount of spectral resources in order to transmit all the user signals and beamforming weights. Consequently, even though certain degradation is expected with respect to the on-ground operation (i.e. beam generation, precoding and detection are done in the terrestrial segment), in the present work we propose to optimize the on-board beam generation process so that the achievable rates do not severally decrease due to the on-board beam generation and the feeder link traffic is kept low.

Concretely, this paper focuses on obtaining an optimal on-board beam generation when linear minimum mean square error (LMMSE) precoding technique in the forward link and  LMMSE detection procedure in the return link are used as  interference mitigation techniques.  This study foresees the presence of a non-channel-adaptive (fixed) on-board beam processing scheme in order to keep payload complexity low. Thus, the problem becomes more difficult in the presence of this fixed process in the payload. In order to deal with this problem, we use a robust optimization framework so that a fixed beam generation can be obtained despite user link channel variation. 

Furthermore, the design for both the forward and return links results the same, which makes it appropriate for the future multibeam satellite systems since it is expected that the same reflector is employed at the return and forward links. Note that the variability of the channel is mainly due to the change of position of the users in consecutive time instants. Numerical simulations show the benefit of our method, which in some scenarios can increase the spectral efficiency over the 6\% and 15\% for return and forward links, respectively, if the DVB-S2 and DVB-RCS modulation and coding parameters (modcods) are used.

To the best of the authors knowledge, this is the first time the problem of on-board beam generation process is treated not only in the forward but also in the return link. In contrast to our preliminar work \cite{design}, where only the forward link was examined, in this paper we focus our attention to the joint forward and return link optimization. {\color{black} In addition, a novel and better robust design is presented based on a tighter upper bound of the optimization problem. This new scheme is conceived considering a first order perturbation approach.} Finally, several detailed evaluations are presented that validate our contribution in detail.
 
The rest of the paper is organized as follows: Section II presents the signal model. A brief introduction of the beam generation process  and the problem characteristics are described in section III. Section IV presents the novel solution  that  the paper proposes. Section V presents a novel robust scheme based on a first order perturbation analysis. Section VI contains a summary of the simulation  results, and eventually the conclusions are given in section VI. 

\textbf{Notation}: Throughout this paper, the following notations are adopted. Boldface upper-case letters denote matrices and boldface lower-case letters refer to column vectors. $ (.)^H$, $(.)^T $, $(.)^*$  and $ (.)^+$  denote a Hermitian transpose, transpose, conjugate  and diagonal (with positive diagonal elements ) matrix, respectively.  $\mathbf{I}_N$ builds $ N\times N$ identity matrix and   $\mathbf{0}_{K\times N}$ refers to an all-zero matrix of size  $ K\times N$. If $\mathbf{A} $ is a $ N\times N$ matrix,   $\mathbf{A}_{1:K}$ refers to taking the  $K$  first rows of the matrix $\mathbf{A}$. $(\mathbf{X})_{ij}$ represents the ($i$-th, $j$-th) element of matrix $\mathbf{X}$. If $\mathbf{B}$ is a $N \times N$ matrix, $\mathbf{A} \leq \mathbf{B} $ implies $\mathbf{A}-\mathbf{B}$ is semidefinite negative.   $\mathbf{a}\prec\mathbf{b}$ means vector $\mathbf{a}$  majorizes vector $\mathbf{b}$.  Finally, $\mathrm E \{\mathbf{.}\}$ and $||.||$ refer to the expected value  operator and the Frobenius norm, respectively.$\cdot$ denotes the matrix Hadamart product.

\section{Signal Model}

Let us consider  a  multibeam satellite communication system, where a single geosynchronous  satellite with multibeam coverage provides fixed broadband services to a large set of  users. To this end,  the satellite is equipped with an array fed reflector antenna whose number of feeds is denoted by $N$. The coverage area is divided into $K$ beams, with 
\begin{equation}
K < N, 
\end{equation}
and the users are assumed to be uniformly distributed within the beams.  By employing a time division multiplexing access (TDMA) scheme, at each time instant the  gateway is serving  a total of $K$ single antenna users (i.e. exactly one user per beam), and it is transmitting (receiving)  information to (from) the same number of the users through the satellite in the forward (return) link. {\color{black} Note that in return link satellite communications generally operate in a multi-frequency TDMA (MF-TDMA) so that different users of the same beam might be allocated to different sub-bands. For the sake of simplicity and without loss of generality, the rest of the paper considers TDMA for the return link. Remarkably, the conceived technique can be accommodated to the multi-band communication by replicating the linear processing at each band due to the frequency flatness of the channel response.}
 
The satellite is assumed to linearly convert a set of $N$ on-board feed signals into the $K$ feeder link signals which are transmitted to the gateway in a frequency multiplexed fashion. Reciprocally, in the forward link, the same linear processing strategy is used to construct the $N$ feed signals from  the $K$ feeder link signals.

\begin{figure}[h!]
\centering
\includegraphics[scale=0.5]{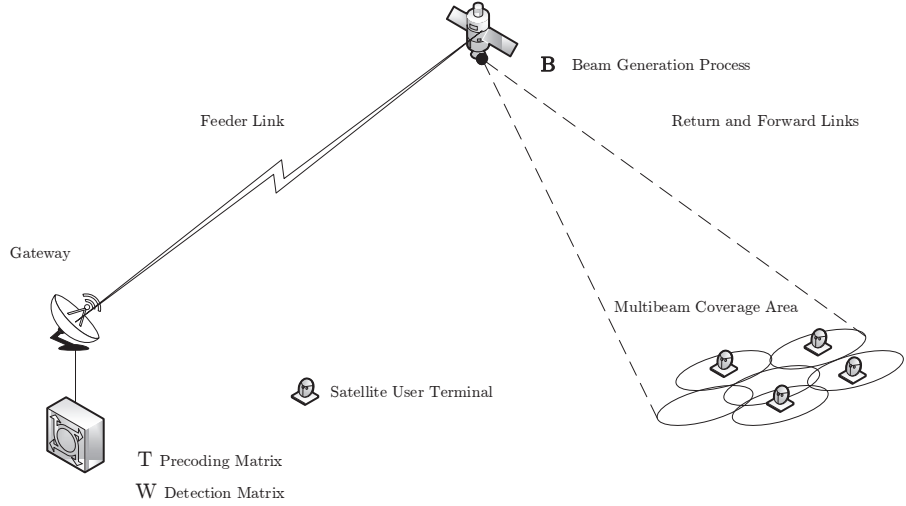}
\caption{Multibeam satellite system with on-board beam generation process. The precoding and detection procedures are done on ground. On the contrary, the beam generation process is carried out at the payload and it is assumed to be constant and the same for forward and return links.}
 \label{Fig22}
\end{figure}

Moreover, since a high throughput system is targeted,  full frequency reuse among beams is assumed so that all beams can share the same frequency resources.  The user link is the communication bottleneck of the whole system. The feeder link is assumed perfectly calibrated and noiseless. Figure \ref{Fig22} summarizes the transmission block diagram.

In the following, the mathematical expressions of the signal model in both the return and forward links are described.

\subsection{Return Link}
As stated above, $K$ denotes the number of users and $N$ is the number of on-board feeds.  Then, the corresponding received signal at the gateway can be modelled as
\begin{equation}
\mathbf{y}_{\text{RL}}=\sqrt{\beta}\mathbf{BHs}+\mathbf{Bn},
\label{ali}
\end{equation}
where  $\mathbf{y}_{RL}=[y_{RL,1}, ...,y_{RL,K}]^{T}$ is a $K\times1$  vector containing the stack of received signals at the gateway. The $K\times 1$ vector $\mathbf{s}$ is the stack of the transmitted independent signals by all users such that $\mathrm E \{\mathbf{s}\mathbf{s}^H\}=\mathbf{I}_K $. Note that, throughout the paper the subscript $_{\text{RL}}$ is used to refer the return link while $_{\text{FL}}$ will denote the forward link. The constant $\beta$ denotes Equivalent Isotropic Radiated Power (EIRP), which is referred to the user terminal transmit power and we assume to be the same for all the users.

In order to radiate the multibeam pattern, the satellite payload is equipped with a smart antenna system (generally an array fed reflector) coined as on-board beam generation process. This system constructs the beam pattern for transmitting and receiving data from the coverage area. Mathematically, the effect of this beamforming appears as the rectangular $K\times N$ fat matrix $\mathbf {B}$. 

The $N\times1$ vector $\mathbf{n}$ accounts for the zero mean Additive White Gaussian Noise (AWGN). We assume unit variance Gaussian noise samples such that 
\begin{equation}
\mathrm E \{\mathbf{Bn}\mathbf{n}^H\mathbf{B}^H\}=\mathbf{BB}^H.
\end{equation}
For radio-frequency design convenience, we will assume that $\mathbf{B}$ is orthonormal so that the feed signals are decoupled at the payload $\left(\mathbf{B}\mathbf{B}^H = \mathbf{I}_K \right)$. Matrix $\mathbf{H}$ is the overall $N\times K$ user link  channel matrix whose element $\text{h}_{ij}$ presents the aggregate gain of the link between the $i$-th satellite feed and the $j$-th user (in the $j$-th beam). This channel  can be decomposed  as follows:
\begin{equation}\label{channel dec}
\mathbf{H} =\mathbf{GD},
\end{equation}
where:
\begin{itemize}
 \item $\mathbf{G}$ is a $N\times K$ matrix that models the feed radiation pattern, the on-board attenuation and path losses. It is  responsible for the interference among users. We assume the elements of $\mathbf{G}$ are normalized so that they have unit variance. 
\item $\mathbf{D}$ is assumed to be a $K\times K$ diagonal matrix which takes into account the atmospheric fading in the user link. 
\end{itemize}
Let us describe more exhaustively matrix $\mathbf{G}$. Its $(k,n)$-th entry can be described as follows
\begin{equation}
\left(G \right)_{k,n} = \frac{G_Ra_{kn}}{4 \pi \frac{d_{k}}{\lambda} \sqrt{K_BT_RB_W}},
\end{equation}
with $d_{k}$ the distance between the $k$-th user terminal and the satellite. $\lambda$ is the carrier wavelength, $K_B$ is the Boltzmann constant, $B_W$ is the carrier bandwidth, $G_R^2$ the user terminal receive antenna gain, and $T_R$ the receiver noise temperature. The term $a_{kn}$ refers to the gain from the $n$-th feed to the $k$-th user. It is important to mention that the $\mathbf{G}$ matrix has been normalized to the receiver noise term. The reader can refer to \cite{Devillers2011} for a more detailed description of the channel model.  

\subsection{Forward Link}
 Analogously to the return link, the signal model of the forward link becomes
\begin{equation}\label{forward_signal}
\mathbf{y}_{\text{FL}}=\gamma\mathbf{H}^{T}\mathbf{B}^{T}\mathbf{x}+\mathbf{w},
\end{equation}
where $K\times1$ vector  $\mathbf{y}_{\text{FL}}$ is  the stack of received signals at each user terminal, and  $\mathbf{x}$ is a  $K\times1$ vector that contains the stack of transmitted symbols. Remarkably, in general wireless communication systems, the channel reciprocity does not hold as uplink and downlink operate in disjoint frequency bands. However, considering our channel modelling {\color{black}, the channel matrix in the forward link differs from the return link in the path loss, feed gain and atmospheric fading. As a result, a scaling factor $\gamma$ can model the different frequency operations. As explained later on, this rescaling factor does not influence the proposed optimization and; therefore, it can be set $\gamma=1$.}

Similarly as in the return link, $\mathbf{w}$ is a $K\times 1$ vector that represents the independent and identically distributed zero mean Gaussian random noise with unit variance such that
\begin{equation} \label{noise_FL}
E \{\mathbf{w}\mathbf{w}^H\}=\mathbf{I}_K.
\end{equation}
Evidently, $\mathbf{B}$ does not influence in the forward link noise covariance matrix. We assume the following average available power constraint:
\begin{equation} \label{Power_FL}
\text{trace}(\mathbf{x}\mathbf{x}^{H})\leq P_{FL},
\end{equation}
where $P_{\text{FL}}$ denotes the total transmit power in the forward link. {\color{black} Note that the transmit power constraint is set without considering the beam generation process $\mathbf{B}$. This is because the power allocation mechanism is located before the array fed reflector system. In addition, it is assumed that the feeds can share the available transmit power. This can be implemented with flexible power amplifiers as described in \cite{multi07}. }

Now, we proceed to jointly optimize matrix $\mathbf{B}$ so that the overall system performance is improved. It is important to remark that $\mathbf{B}$ must be the same for both the optimization of the return and forward links in order to reduce the payload cost. In addition, this matrix needs to be constant in order to keep the payload complexity low and minimize the feeder link spectral resources.

\section{Problem Formulation}

Let us assume that the gateway has perfect Channel State Information (CSI) and uses LMMSE as described in \cite{Peel2005} for precoding in the forward link and LMMSE filtering for multiuser detection in the return link. These techniques have been pointed out as efficient methods due to both its interference rejection capabilities and fairness among beams while preserving a low computational complexity \cite{Arnau2011}. 

This work resorts to the minimization of the trace of the MSE matrix both at the forward and return links that results from the use of LMMSE precoding and detection. Let us briefly outline the overall mathematical derivation: 

\begin{enumerate}
  \item First, the MSE matrix of the return link is computed assuming LMMSE detection.
  \item Second, the MSE matrix of the forward link is computed assuming LMMSE precoding. 
  \item Third, an upper bound of the MSE minimization in the return link is presented.
  \item Finally, a novel robust beam generation process in the return link, which considers the aforementioned upper bound is obtained. For the forward link case, the optimal design yields to the same solution as it is described.
\end{enumerate}

Remarkably, the design of the optimal $\mathbf{B}$ is imposed to be non channel dependent. We show that the optimal $\mathbf{B}$ in the forward and return links results to be the same; thus, fulfilling one of the constraints of the system.

\subsection{Return Link}
As a first step, let us define $\mathbf{W}^H$ as the LMMSE filter that detects $K$ received signals  at the gateway such that $\hat{\textbf{s}}=\textbf{W}^{H}\textbf{y}_{\text{RL}}$; composed by  $\hat{ \text{s}}_{i}$ which denotes  the $i$-th element of the detected signal (for $i$-the user) in the gateway. In this context, the MSE of $i$-th user is achieved as follows 
 \begin{equation}
\text{MSE}_{\text{RL},i}=\mathrm E \{|\text{s}_{i}-\hat{ \text{s}}_{i}|^2\},
\end{equation} 
where $\text{s}_{i}$  represents the $i$-th element of transmit signal  vector  (for $i$-the user) for a total of $K$ users such that  $\textbf{s}=(s_{1},..., s_{K})^{T}$.  

It is well known that the mathematical expression of LMMSE filter becomes 
\begin{equation}\label{filter}
\mathbf{W}^H= \left(\mathbf{I}_K+ \beta\mathbf{H}^{H}\mathbf{B}^{H}\mathbf{BH} \right)^{-1}\mathbf{H}^{H}\mathbf{B}^{H},
\end{equation}
and the MSE matrix after the use of this filter is
\begin{equation}\label{MSE}
\textbf{MSE}_{\text{RL}}=\left(\mathbf{I}_{K}+\beta\mathbf{H}^{H}\mathbf{B}^{H}\left(\mathbf{B}\mathbf{B}^{H}\right)^{-1}\mathbf{B}\mathbf{H}\right)^{-1}.
\end{equation}

Without loss of generality, we restrict  $\mathbf{B}$ to be orthonormal such that $\mathbf{B}\mathbf{B}^{H}=\mathbf{I}_K$.  The sum of MSE in the return link is defined as
\begin{equation}\label{MSE1}
\text{SMSE}_{\text{RL}}=\text{trace}\left(\left(\mathbf{I}_{K}+\beta\mathbf{H}^{H}\mathbf{B}^{H}\mathbf{B}\mathbf{H}\right)^{-1}\right).
\end{equation}

Now, let us assume for a moment that $\mathbf{B}$ can be channel adaptive (i.e the payload can modify $\mathbf{B}$ depending on the channel variations) . Then, the corresponding problem is formulated  as\\
\begin{equation}\label{1}
\min_{\mathbf{B}} \quad \text{trace}\left(\left(\mathbf{I}_{K}+\beta\mathbf{H}^{H}\mathbf{B}^{H}\mathbf{B}\mathbf{H}\right)^{-1}\right)
\end{equation}
$~~~~~~~~~~~~~~~~~~~~~~s.t.~~~~~~~~~\mathbf{BB}^H=\mathbf{I}_K.~~~~~~~~~~~~~~~$\\

It is important to remark that the authors in \cite{bertrand} showed that the presence  of $\mathbf{B}$ increases the SMSE$_{\text{RL}}$ in the gateway. Mathematically,
\begin{equation}\label{MSE11}
 \text{trace}\left(\left(\mathbf{I}_{K}+\beta\mathbf{H}^{H}\mathbf{B}^{H}\mathbf{B}\mathbf{H}\right)^{-1}\right)\geq \text{trace}\left(\left(\mathbf{I}_K+\beta\mathbf{H}^{H}\mathbf{H}\right)^{-1}\right).
\end{equation}
 Indeed, in \cite{bertrand} it was shown that with the following Singular Value Decomposition (SVD) of the channel  $\mathbf{H}={\mathbf{U} } \boldsymbol{\Phi}{\mathbf{V}}^H $, an optimal design of  $\mathbf{B}$ can  be worked out as  \begin{equation}\label{design B}
\mathbf{B}={\mathbf{U}}_{1:K}^H,
\end{equation}
where  $\mathbf{U}_{1:K}^{H}$ denotes the $K$ first rows of the matrix $\mathbf{U}^{H}$. In fact, it can be easily seen that this particular solution reaches equality in \eqref{MSE11} and; thus, minimizes the SMSE$_\text{{RL}}$.

In the present work,  $\mathbf{B}$  is assumed to be non-channel adaptive, therefore, the design of  $\mathbf{B}$ in \eqref{design B} cannot be considered.  Even though the channel appears to be variable at each realization, we aim at finding the best possible non-channel adaptive design of $\mathbf{B}$.  In this context, let us decompose the channel as follows
\begin{equation}\label{channel model}
\mathbf{H}\triangleq\bar{\mathbf{H}}+ \boldsymbol{\Delta},
 \end{equation}
where:
\begin{itemize}
\item  ${\bar{\mathbf{H}} }$ represents the mean value of the channel.
\item  $ \boldsymbol{\Delta} $ models the difference between the actual value of the channel and its mean. It indicates the variability of the channel in consecutive time instants as already explained in section I. 
\end{itemize}
We assume that the actual channel $\mathbf{H}$ lies in the neighbourhood of a nominal channel $\bar{\mathbf{H}}$ that is known to the gateway. In particular, we consider that $\mathbf{H}$ belongs to the uncertainty region $ \mathcal{H} \triangleq \{ \mathbf{H}: ||\mathbf{H}-\bar{\mathbf{H}}||\leq \alpha\}$ which is an sphere centered at $\bar{\mathbf{H}}$ with the radius $\alpha$.

Interestingly, the channel  model in \eqref{channel model} resembles the modeling of a MIMO system with imperfect CSI at the transmitter which has been solved as a worst case optimization problem in \cite{Wang2009,palomar,pas06}. With this perspective for the return link, the worst case robust design is proposed, which leads to a maximin or minimax formulation:
\begin{equation}\label{worst}
\min_{\mathbf{B}}~~~\max_{{\Delta}}~~~~\text{trace}\left(\left(\mathbf{I}_{K}+\beta\mathbf{H}^{H}\mathbf{B}^{H}\mathbf{B}\mathbf{H}\right)^{-1}\right)
 \end{equation}
$~~~~~~~~~~~~~~~~~~~~~~~~~~s.t.~~~~~~~~~\mathbf{BB}^H=\mathbf{I}_K.~~~~~~~~~~~~~~~~~$\\
Prior to obtaining the solution of \eqref{worst}, let us focus on the forward link optimization problem, which is similarly derived.

\subsection{Forward Link}
In the forward link, the zero forcing precoding with a regularized inversion is assumed \cite{Peel2005}. In this case, the linear precoding is expressed as
\begin{equation} \label{precoding}
\mathbf{x}=\mathbf{Tc},
\end{equation}
where $\mathbf{T}$ is the $K\times K$ precoding matrix at the gateway and $\mathbf{c}$ is the $K\times 1$ transmit symbol vector at all feeds such that $\mathrm E \{\mathbf{c}\mathbf{c}^H\}=\mathbf{I}_K $.  In this context, the corresponding precoding matrix $\mathbf{T}$ is expressed as
 \begin{equation}\label{precoder}
\mathbf{T}=\sqrt{\rho} \mathbf{B}^*\mathbf{H}^{*}\left(\frac{K}{P_{\text{FL}}}\mathbf{I}_K+\mathbf{H}^{T}\mathbf{B}^{T}\mathbf{B}^{*}\mathbf{H}^{*}\right)^{-1},
 \end{equation}
where the value of the constant $\rho$ has  to  comply with the forward link power constraint as follows
\begin{equation}
\text{trace}\left(\mathbf{TT}^{H}\right) \leq P_{\text{FL}}.
\label{VGD_F}
\end{equation}
This particular kind of precoder is used to find an optimal balance between achieving  signal gain and limiting the multiuser  interference. Similar to the return link,  MSE$_{\text{FL},i}$ is defined as 
\begin{equation}
\text{MSE}_{FL,i}=\mathrm E \{|\text{c}_{i}-\hat{ \text{c}}_{i}|^2\},
\end{equation}
where $\text{MSE}_{\text{FL},i}$ refers to the MSE received by $i$-th user. Similarly, $\textbf{c}=(c_{1},..., c_{K})^{T}$ and $\hat{\textbf{c}}=(\sqrt{\rho})^{-1}\mathbf{y}_{\text{FL}}=(\hat{c}_{1},..., \hat{c}_{K})^{T}$ are the transmitted  and  received signals for $K$ users, respectively. In this context, $\text{c}_{i}$  represents the transmitted signal for $i$-the user and   $\hat{ \text{c}}_{i}$ denotes  the signal received by user $i$-th.
The MSE matrix in the forward link  can be calculated as follows
\begin{equation}
\textbf{MSE}_{\text{FL}}=\mathrm E \left\{\left((\sqrt{\rho})^{-1}\mathbf{y}_{\text{FL}}- \textbf{c}\right)\left((\sqrt{\rho})^{-1}\mathbf{y}_{\text{FL}}- \textbf{c}\right)^{H} \right\},
\end{equation}
which can be rewritten as in \eqref{vasat11}. 
\begin{equation}
\textbf{MSE}_{\text{FL}}=\frac{K}{P_{\text{FL}}}\left( \left(\mathbf{H}^{T}\mathbf{B}^{T}\mathbf{B}^{*}\mathbf{B}^{T}\mathbf{B}^{*}\mathbf{H}^{*}+\frac{K}{P_{\text{FL}}}\mathbf{I}_K\right)\left(\mathbf{H}^{T}\mathbf{B}^{T}\mathbf{B}^{*}\mathbf{H}^{*}+\frac{K}{P_{\text{FL}}}\mathbf{I}_K \right)^{-2} \right).
\label{vasat11}
\end{equation}
As in the return link, we concentrate our efforts to minimize the sum of MSE, this is
\begin{equation}
\text{SMSE}_{\text{FL}} = \text{trace}(\textbf{MSE}_{\text{FL}}),
\label{WER1222}
\end{equation}
where, recalling that $\mathbf{BB}^{H}=\mathbf{I}_K$ and we consider the following property, $\text{trace}(\mathbf{A})=\text{trace}(\mathbf{A}^{T})$ where $\mathbf{A}$ is a square matrix, then we have that
\begin{equation}
\text{SMSE}_{FL}= \frac{K}{P_{\text{FL}}}\text{trace}\left(\left(\mathbf{H}^{H}\mathbf{B}^{H}\mathbf{B}\mathbf{H}+\frac{K}{P_{\text{FL}}}\mathbf{I}_K\right)^{-1}\right).
\label{DSA}
\end{equation}
The worst case optimization problem thanks to the channel decomposition in \eqref{channel model} can be formulated as follows
\begin{equation}\label{worst2}
\min_{\mathbf{B}}~~~\max_{{\Delta}}~~~~\text{trace}\left(\left(\mathbf{H}^{H}\mathbf{B}^{H}\mathbf{B}\mathbf{H}+\frac{K}{P_{\text{FL}}}\mathbf{I}_K\right)^{-1}\right)
 \end{equation}
$~~~~~~~~~~~~~~~~~~~~~~~s.t.~~~~~~~~~\mathbf{BB}^H=\mathbf{I}_K.~~~~~~~~~~~~~~~~$\\

Note that the return link optimization \eqref{worst} and the forward link one \eqref{worst2} are the same except for a scalar value. In next section we show that both lead to the same optimal design; thus confirming a natural uplink downlink physical duality.

\section{$\mathbf{B}$ optimization}

This section tackles with the main objective of this paper. An optimally designed $\textbf{B}$ for problems \eqref{worst} and \eqref{worst2} is  presented.  Two main steps are followed. The first step provides a brief description of an upper bound for the SMSE. The second step  proposes a  design for $\mathbf{B}$ such that it minimizes the proposed SMSE upper-bound obtained in the first step. The design is done for the return link and extended to the forward link.

Prior to presenting the optimal design, we need to introduce the next lemma.

\textbf{Lemma 1}. \emph{Assuming an arbitrary square matrix $\mathbf{A}$, the next equation holds}
\begin{equation}
\text{trace}\left(\left(\mathbf{I}_{K}+\mathbf{AA}^{H}\right)^{-1}\right)=\text{trace}\left(\left(\mathbf{I}_{K}+\mathbf{A}^{H}\mathbf{A}\right)^{-1}\right).
\end{equation}
\begin{proof}
It is a direct consequence of inversion matrix lemma.
\end{proof}
By considering $\mathbf{A} = \sqrt{\beta}\mathbf{BH}$,  the SMSE$_{\text{RL}}$ in problem \eqref{worst} can be rewritten as
\begin{equation}\label{re-mse}
\text{trace}\left(\left(\mathbf {I}_K + \beta\mathbf {B}\mathbf{Z}\mathbf {B}^{H}\right)^{-1}\right),
\end{equation}
where $\mathbf{Z}=\mathbf {H}{\mathbf {H}}^H=\bar{\mathbf {H}}\bar{\mathbf {H}}^H+\bar{\mathbf {H}}\boldsymbol{\Delta}^{H}+\boldsymbol{{\Delta}}\bar{\mathbf {H}}^{H}+\boldsymbol{{\Delta}{\Delta}}^H$ is a $N\times N$  matrix. 
We propose an upper bound of SMSE$_{\text{RL}}$ as follows

\textbf{Theorem 1:}\emph{ The SMSE$_{\text{RL}}$ is upper bounded by 
\begin{equation}\label{compare} 
 \text{trace}\left(\left(\mathbf {I}_K +\beta\mathbf {B}\mathbf{Z}\mathbf {B}^{H}\right)^{-1}\right) \leq \text{trace}\left(\left(\mathbf{I}_K+\beta\mathbf{B}\breve{\mathbf{Z}}\mathbf{B}^H\right)^{-1}\right)
\end{equation}
where
\begin{equation}\label{222}
\breve{\mathbf{Z}}\triangleq \bar {\mathbf{U} }(\boldsymbol{\bar {\Sigma}}-\epsilon_{H} {\mathbf{I}_N})^{+}\bar {\mathbf{U}}^{H},
\end{equation}
so that $\bar{\mathbf{H}}\bar{\mathbf{H}}^{H}=\bar {\mathbf{U} }\bar{\boldsymbol{ {\Sigma}}}\bar{\mathbf{U} }^{H}$ is the eigen-decomposition of matrix $\bar{\mathbf{H}}\bar{\mathbf{H}}^{H}$.
The scalar value $\epsilon_{H}$ is defined as 
\begin{equation}\label{intcos}
 \epsilon_{H}\triangleq  2 \alpha \delta_{max}(\bar {\mathbf{H}})
\end{equation}
where $\delta_{max}(\mathbf{C})$ denotes the maximum singular value of $\mathbf{C}$  matrix.}
\begin{proof}
See \cite[sec.7.3.1]{palomar}. 
\end{proof}
As a result,  a worst-case  $\text{SMSE}_{\text{RL}}$  can be obtained in practice by using the lower bound $\breve{\mathbf{Z}}$ in lieu of ${\mathbf{Z}}$. However, it is important to mention that some values of $\alpha$ lead to infeasible MSE$_{\text{RL}}$ solutions, that is, for a large value of $\alpha$ the matrix  \eqref{222} might become {\color{black} low rank since $()^{+}$ operator delivers 0 whenever the diagonal entry is nonpositive}.  In order to avoid this circumstance, the value of $\alpha$ has to be checked and, if necessary, decreased so that the feasibility condition of the problem \eqref{worst} is hold. 

In order to obtain a robust design, the target is to minimize the proposed upper-bound of SMSE$_{RL}$  in \eqref{compare} instead of \eqref{worst}. In this case, the corresponding problem  is formulated as

\begin{equation}\label{2}
\min_{\mathbf{B}}~~~~~~~\text{trace}\left(\left(\mathbf{I}_K+\beta\mathbf{ B} \breve{\mathbf{Z}}\mathbf{B}^{H}\right)^{-1}\right)
\end{equation}
$~~~~~~~~~~~~~~~~~~~~~~~~~~s.t.~~~~~~~~~~~\mathbf{BB}^H=\mathbf{I}_K.~~~~~~~~$\\

The solution to this optimization problem is described in the next theorem.

\textbf{Theorem 2:} \emph{ Let $\mathbf{B}$ and $\bar{\mathbf{L}}^{H}$ be two matrices of size $K\times N$ and $N \times N$, respectively. Then, the upper bound of SMSE is minimized if $\mathbf{B}$ is selected as the first K rows of the matrix $\bar{\mathbf{L}}^{H}$, that is 
\begin{equation}\mathbf{B}^{\star}=\bar{\mathbf{L}}_{1:K}^{H},
\label{design B1}
\end{equation}
where $\mathbf{B}^{\star}$ denotes the optimal design of $\mathbf{B}$.}
\begin{proof}
See Appendix A.
\end{proof}
{\color{black}
\textbf{Remark:} It is important to mention that the derivation of theorem 2 differs to theorem 1 in \cite{1223549}. The main difference relays on the constraint since in \cite{1223549} a total power constraint is considered
\begin{equation}
\text{trace}\left(\mathbf{B}\mathbf{B}^H \right) \leq P,
\end{equation}
where as  this paper assumes
\begin{equation}
\mathbf{BB}^H=\mathbf{I}_K,
\end{equation}
which involves further mathematical developments as described in Appendix A.}

Before starting with the forward link case, let us remark that $\mathbf{B}^{\star}$ only needs statistical channel knowledge in order to be computed. Moreover, its design does not depend on $\alpha$. Indeed, the value of $\alpha$ affects only on the resulting SMSE$_{\text{RL}}$. This is due to the optimization of an upper bound of the problem instead of the problem itself. Now, let us  proceed with the forward link optimization.

In the forward link the optimization problem can be formulated as follows
\begin{equation}\label{worst3}
\min_{\mathbf{B}}~~~~~~~{\text{trace}\left(\left(\mathbf {B}\breve{\mathbf{Z}}\mathbf {B}^{H}+\frac{K}{P_{\text{FL}}}\mathbf{I}_K\right)^{-1}\right)}
\end{equation}
$~~~~~~~~~~~~~~~~~~~~~~~~~s.t.~~~~~~~~~~~\mathbf{BB}^H=\mathbf{I}_K.~~~~~~~~~~~~~~~~~~~~$\\

In can be observed that the  optimal solution of \eqref{worst3} is \eqref{design B1}. The sketch of the proof is similar to the one presented previously for the return link and; thus, we only comment it. The idea is to check whether the term $\frac{K}{P_{\text{FL}}}$ does not influence the optimal value of \eqref{worst3} which can be easily observed in appendix A. {\color{black} Consequently, the scaling factor due to the channel variations $\gamma$ does not influence the optimization, either.} Remarkably, this derivation is different from the one presented in our preliminary work in \cite{design}, because this paper so encompasses the forward and return link optimizations.

{\color{black}

Note that the robust beamforming design has the same eigenvectors as the nominal channel matrix $\bar{\mathbf{H}}\bar{\mathbf{H}}^H$. In other words, the presented robust design only considers eigenvalue variations due to the different user positions. In the next section, the impact on the eigenvectors is analysed.

\section{ First Order Perturbation Analysis}
As discussed in the previous sections, the underlying optimization problem \eqref{worst2} shall be lower bounded in order to obtain a closed-form solution. This is done by means of considering upper bounds of $\mathbf{Z}$. 

Indeed, the proposed perturbation model can be described as
\begin{equation}
\mathbf{Z} = \left( \bar{\mathbf{U}}_s + \Delta\mathbf{U}_s \right) \left(\bar{\boldsymbol{\Sigma}}_s + \Delta\boldsymbol{\Sigma}_s \right) \left(\bar{\mathbf{U}}_s + \Delta\mathbf{U}_s \right)^H + \left(\bar{\mathbf{U}}_n + \Delta\mathbf{U}_n \right) \left(\bar{\boldsymbol{\Sigma}}_n + \Delta\boldsymbol{\Sigma}_n \right) \left(\bar{\mathbf{U}}_n + \Delta\mathbf{U}_n \right)^H,
\end{equation}
where the $\mathbf{U}$ denotes the matrix containing the eigenvectors and $\boldsymbol{\Sigma}$ is a diagonal matrix which contains the eigenvalues. Subindex $s$ denotes the non-zero signal space whereas $n$ the signal space that is spanned by the zero valued eigenvalues (i.e. the null space of $\mathbf{Z}$). All $\Delta\mathbf{U}_s$, $\Delta\boldsymbol{\Sigma}_s, \Delta\mathbf{U}_n,\Delta\boldsymbol{\Sigma}_n$ are generated by a perturbed version of $\bar{\mathbf{Z}}$:
\begin{equation}
\mathbf{Z} = \bar{\mathbf{Z}} + \Delta\mathbf{Z},
\end{equation}
where 
\begin{equation}
\bar{\mathbf{Z}} = \bar{\mathbf{H}}\bar{\mathbf{H}}^H,
\end{equation}
and
\begin{equation}
\Delta\mathbf{Z} = \bar{\mathbf{H}}\Delta^H + \Delta\bar{\mathbf{H}}^H + \Delta\Delta^H.
\end{equation}
Under this context, $\bar{\mathbf{U}}$ denotes the eigenvector of the nominal matrix $\bar{\mathbf{Z}}$ whereas $\bar{\boldsymbol{\Sigma}}$ a matrix containing its eigenvalues. The other matrices with the $\Delta\cdot$ prefix denote the corresponding perturbation matrices.

The previous section has implicitly considered two assumptions. First, it has been assumed that the channel variations do not modify the dimension of the null space so that $\Delta\boldsymbol{\Sigma}_n$ remains as a zero matrix. Second, it has been assumed that $\Delta\mathbf{U}_s = 0$ ,which might not be true in certain cases \cite{liu8}. The aim of this section is to consider the effect of this later perturbation in order to obtain a tighter upper bound of  $\mathbf{Z}$ than the presented in the previous section. Remarkably, the following inequality holds
\begin{equation}
\mathbf{Z} \geq \widehat{\mathbf{Z}} \geq \breve{\mathbf{Z}},
\end{equation}
where $\breve{\mathbf{Z}}$ only considers perturbations at the eigenvalues whereas $\widehat{\mathbf{Z}}$ considers both perturbations at both eigenvalues and  eigenvectors ($\Delta\mathbf{U}_s$). Next theorem provides an approximate solution whenever these both perturbations are considered.

\textbf{Theorem 3} \emph{The beamforming matrix that optimizes the MSE upper bound when considering both eigenvector and eigenvalue perturbations is}
\begin{equation}\label{cos2}
\widehat{\mathbf{B}^{*}} = \widehat{\mathbf{U}} \left( \boldsymbol{\bar {\Sigma}}-\epsilon_{H} {\mathbf{I}_N} \right)^{+} \widehat{\mathbf{U}}^H,
\end{equation}
\emph{where}
\begin{equation}
\widehat{\mathbf{U}} = \bar{\mathbf{U}}_s + \epsilon_H\bar{\mathbf{U}}_s\widehat{\mathbf{R}} + \epsilon_H\bar{\mathbf{U}}_n\bar{\mathbf{U}}_n^H\bar{\mathbf{U}}_s\bar{\boldsymbol{\Sigma}}_s^{-1},
\end{equation}
\emph{and}
\begin{equation}
\widehat{\mathbf{R}} = \mathbf{D}\cdot \left(\mathbf{U}_s^H\mathbf{U}_s\bar{\boldsymbol{\Sigma}} + \bar{\boldsymbol{\Sigma}}\mathbf{U}_s^H\mathbf{U}_s \right),
\end{equation}
\emph{and the $g,f$-th entry of $\mathbf{D}$ is}
\begin{equation}
\frac{1}{\lambda_f - \lambda_g},
\end{equation}
\emph{for $f\neq g$ and $\lambda_f$ for $f=1,\ldots,N$ denote the eigenvalues of $\bar{\mathbf{H}}\bar{\mathbf{H}}^H$.}
\begin{proof}
See Appendix B.
\end{proof}
Note that for this case, the eigenvectors of the beamforming matrix take a different value from the  nominal matrix. In addition, the larger $\alpha$ the more different are the eigenvectors from the nominal channel matrix ones.

As we have already seen, the beam generation process both on the forward and return links leads to the same matrix $\mathbf{B}$, which is fixed. Now, it is time to compare the benefits of this design in front of the current beam processing deployments. 
}
\section{ Simulation Results}
In order to show the performance of our proposal, this  section presents a numerical evaluation of the conceived technique. Our baseline scenario is  an array fed reflector antenna  and matrix $\mathbf{B}$ that have been provided by ESA in  the framework of a study on next generation multibeam satellite systems. The number of feeds is assumed to be $N=155$  and $K=100$ beams  that are covering the whole Europe area. 

Results have been averaged over a total of 1000 user link channel realizations. Note that, only atmospheric fading due to rain effect is considered in the user link channel and further refinements of the  channel are neglected. This simple characterization is useful for the intended comparisons and it is a general practice in the evaluation of multibeam satellite systems. 

The randomness of the channel is due to the user positions which are assumed  to be uniformly distributed within the beams. {\color{black} In addition, we will assume that each user employs all available spectrum and the atmospheric fading is modelled as in \cite{au97}}. 

Recall that,  full frequency reuse among beams and noiseless feeder link  have been considered in this work. In the sequel, we compute different performance metrics. First, the SINR for each user after employing interference mitigation techniques among  users is presented. Then, with that SINR value,  the throughput is  inferred according to DVB-RCS and DVB-S2 standards for the return and forward links, respectively \cite{DVB,DVB2}.  Furthermore, the simulation results also provide the associated Cumulative Distribution Function (CDF) of SINR  which shows the availability of the user link. In this case, the instantaneous availability indicator for the $k$-th user is given by
\begin{equation}
A_k = g(\text{SINR}_k)
\end{equation}
which is equal to 0 if the user link is unavailable (i.e, if the instantaneous SINR is lower than that required by the lowest modcod for the return link, i.e. $\text{SINR}_k < 1.7$dB, and for the forward link, i.e. $\text{SINR}_k < -2.72$dB ) and is equal to 1 otherwise. We also present the Shannon capacity\footnote{Of course, we refer to the use of the Shannon formula instead of the channel capacity.} obtained from the user SINR,
\begin{equation}
C_{\text{Shannon}} = \log_2(1 + \text{SINR}),
\end{equation}
 and assuming that interference is treated as Gaussian noise. This measurement serves us to see the potential of our work independently of the satellite standard modulations and channel coding both for the forward and return links.

Another performance metric to be considered is the fairness among beams. Note that this is of great interest for satellite operators where near to equal achievable data rates per beam are the target. For this purpose, we present the throughput index of dispersion, defined as
\begin{equation}
\text{Index of Dispersion} = \frac{\sigma_{\text{Th}}}{\mu_{\text{Th}}},
\end{equation} 
where $\sigma_{\text{Th}}$ and $\mu_{\text{Th}}$ correspond to the variance and the mean of the user throughputs, respectively. This metric provides an indicator of how the data rates are dispersed with respect to the mean. The larger the index of dispersion is, the less the fairness the system achieves.

For a best practice, as upper bound for the achievable rates we consider only on-ground processing at the gateway (i.e. no on-board processing) as it is described in \cite{bertrand}. From the return link point of view, the  received signal  \eqref{ali}, which is  based on this on-ground  scenario,  is rewritten  as 
\begin{equation}
\mathbf{y}_{\text{RL}}=\mathbf{T}_{\text{on-ground}}^{H}\left(\mathbf{Hs}+\mathbf{n}\right),
\label{signal222}
\end{equation}
where
\begin{equation}
\mathbf{W}_{\text{on-ground}} = \mathbf{H}\left(\mathbf{H}\mathbf{H}^H + \mathbf{I}_K \right)^{-1}
\end{equation}
denotes the LMMSE detector filter at the gateway. Note that the linear processing is similar to \eqref{filter} but in this case it has been assumed that no beam processing is done. Considering the forward link, the received signal by the user terminals with this on-ground technique can be represented as
\begin{equation}\label{forward_signal222}
\mathbf{y}_{\text{FL}}=\mathbf{H}^{T}\mathbf{T}_{\text{on-ground}}\mathbf{x}+\mathbf{w}.
\end{equation}

It is important to remark that although large data rates can be obtained if all the processing is carried out on ground, the required feeder link spectral resources severally increase, leading to an inefficient system. 
 
To sum up, in order to test the validity of the  derived  theoretical results in section IV, we  compute the spectral efficiency of the following multibeam satellite system using precoding and detection algorithms for forward and return links respectively:

\begin{itemize}
\item $\mathbf{B}$ based on a geographical reasoning (reference).
\item $\mathbf{B}^*$ proposed  by this study in  \eqref{design B1}.
\item $\widehat{\mathbf{B}}^*$ proposed  by this study in  \eqref{cos2}.
\item On ground processing (upper bound). 
\end{itemize}

In the sequel, the results are separated into two different subsections, return and forward link. In this context,  the same fixed optimal design of  on-board beamforming matrix is computed  since this optimal design depends on the right eigen vector of channel average matrix, $\bar{\mathbf{H}}$. This is computed empirically considering the aforementioned 1000 channel user realizations.

\subsection{ Return Link}
\begin{figure}[tb!]
\centering
\includegraphics[scale=0.6]{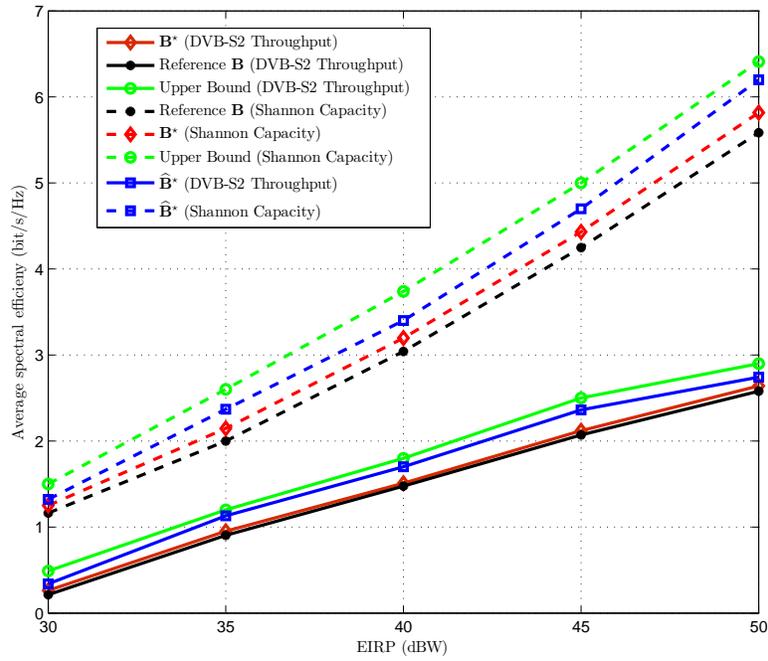}\vspace{0.3cm}
\caption{Return link throughput values over different user EIRP ($\beta$).}
\label{fig1}
\end{figure}

The return link operates at 30GHz, and is based on DVB-RCS standard \cite{DVB} and we target a  Packet Error Rate (PER) of $10^{-7}$. Figure \ref{fig1}  depicts the evolution of the total average throughput (bits/symbol)  as a function of the user EIRP ($\beta$) for different scenarios. Although by means of using the DVB-RCS standard the obtained throughput gain is limited  when the Shannon capacity is considered, higher gains are obtained with respect to the reference scenario .  In other words, other modcods design would improve the benefits of the proposed technique with respect to the reference scenario. {\color{black} Note that the proposed robust design that consider the eigenvector perturbation improves the system throughput with respect to the design that only considers eigenvalue variations.} Indeed, our proposal is approaching the upper bound of the on ground design.

\begin{figure}[tb!]
\centering
\includegraphics[scale=0.6]{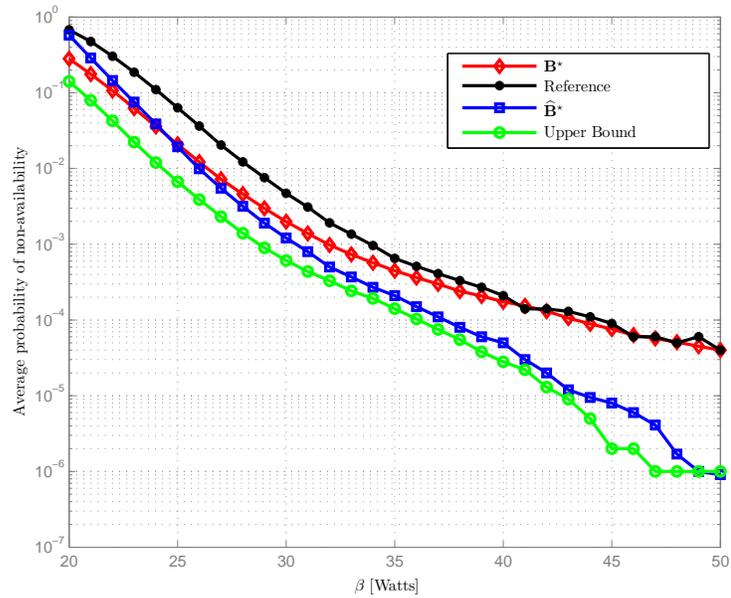}
\caption{Return link availability.}
\label{fig2}
\end{figure}

\begin{figure}[tb!]
\centering
\includegraphics[scale=0.6]{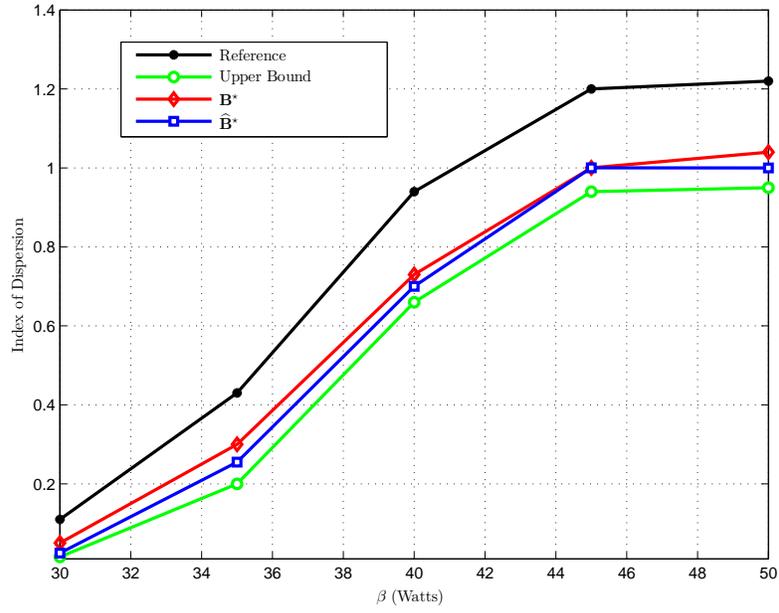}
\caption{Return link throughput index of dispersion.}
\label{fairness_RL}
\end{figure}

\begin{figure}[tb!]
\centering
\includegraphics[scale=0.6]{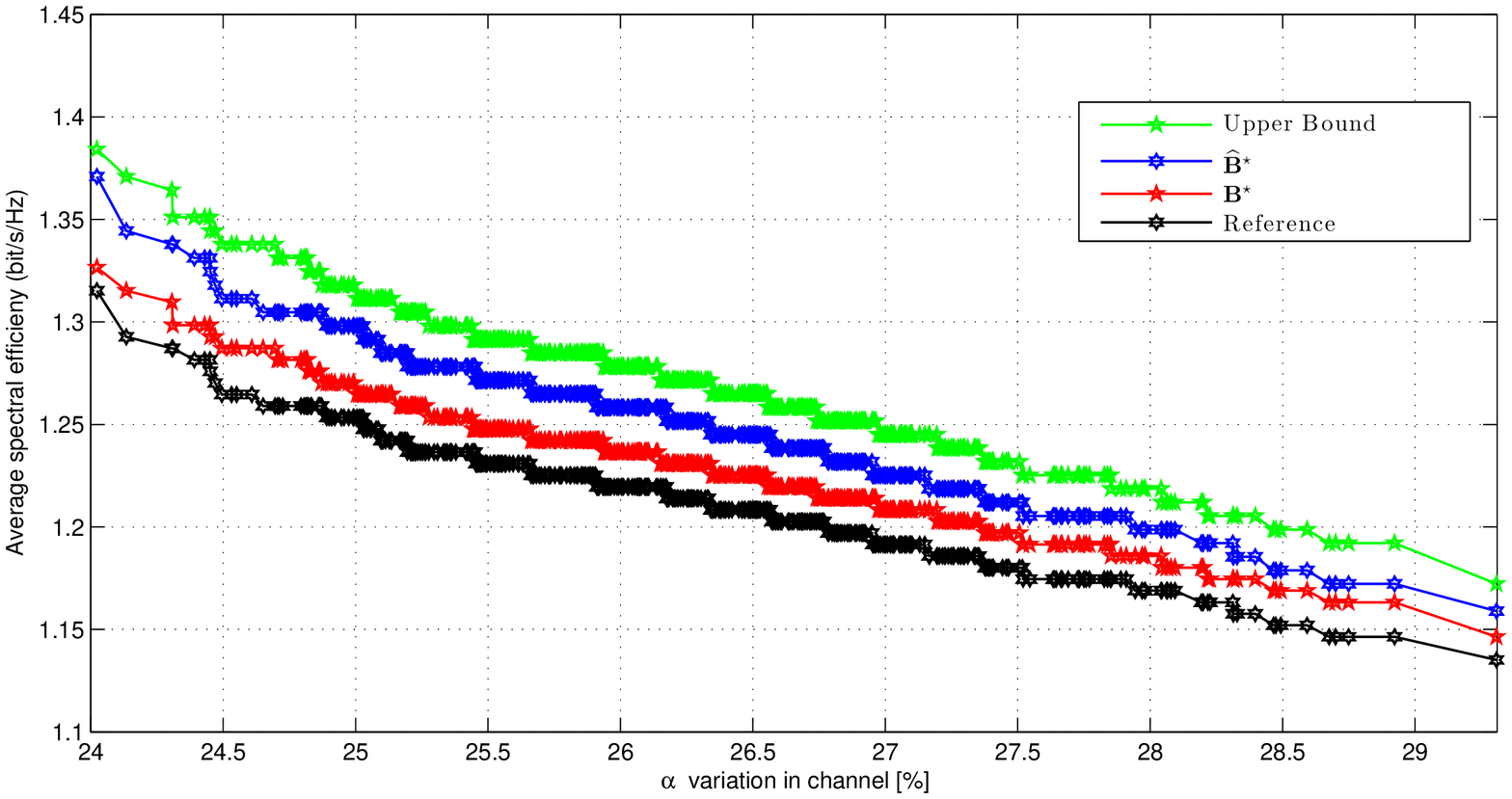}
\caption{Return link throughput with respect to channel variations.}
\label{alpha_RL}
\end{figure}

The corresponding availability probability  is also provided in Figure \ref{fig2}.  In this case, our proposal also improves the reference scenario, leading to an increase of the system availability. Remarkably, the fairness among beams is also improved as it is depicted in Figure \ref{fairness_RL}. Lower values of dispersion index are obtained with our technique with respect to the reference design. 

Finally, we study the impact of the channel variations on the beam processing design. Bearing in mind that $\alpha$ in \eqref{intcos} determines this variation, we compute this value and we present its corresponding average throughput values in Figure \ref{alpha_RL}. The values of $\alpha$ are selected so that the feasibility of MSE$_{\text{RL}}$  in \eqref{222} holds. It implies that  
\begin{equation}\label{vahidjoroughi222}
(\boldsymbol{\bar {\Sigma}}-\epsilon_{H} {\mathbf{I}_N})_{ii}\geq 0~~~~~~~~~~   \forall i=1, ..., N. 
\end{equation}
For a large value of $\alpha$ the  matrix \eqref{vahidjoroughi222} might become semidefinite negative and; thus, changes the nature of the problem. In order to avoid this,  $\alpha$ has to be checked  so that the matrix \eqref{vahidjoroughi222}  always remains semidefinite positive. It is observed that the larger $\alpha $ values, the less the throughput is obtained due to the channel mismatch. 

\subsection{ Forward Link}
The forward  link is assumed to operate at 30GHz and  is  based on DVB-S2 standard with a  PER of $10^{-6}$. Note that the working points were extrapolated from the PER curves reported in the DVB-S2 guidelines document \cite{DVB2}. Based on \cite{DVB2}, it is possible to find a  relationship between the required received SINR and the spectral efficiency achieved by DVB-S2 standard.

\begin{figure}[tb!]
\centering
\includegraphics[scale=0.6]{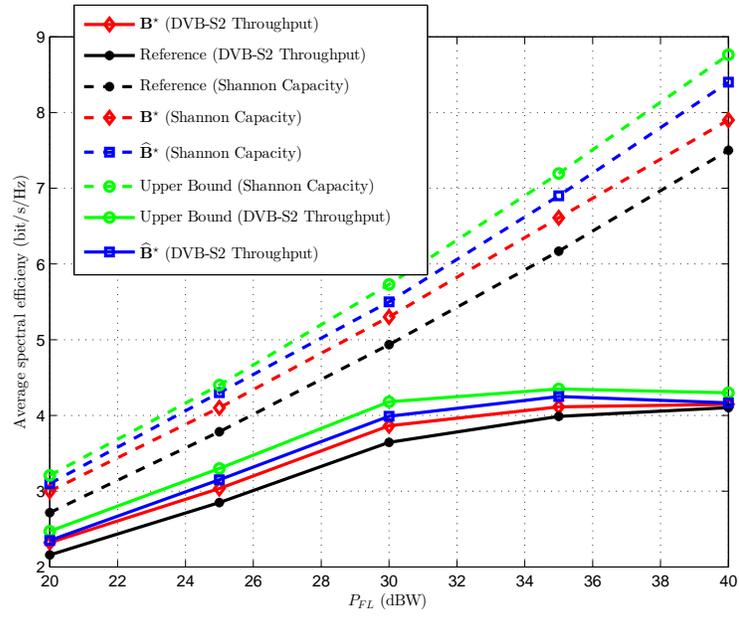}
\caption{Forward link throughput values.}
\label{fig3}
\end{figure}

\begin{figure}[tb!]
\centering
\includegraphics[scale=0.6]{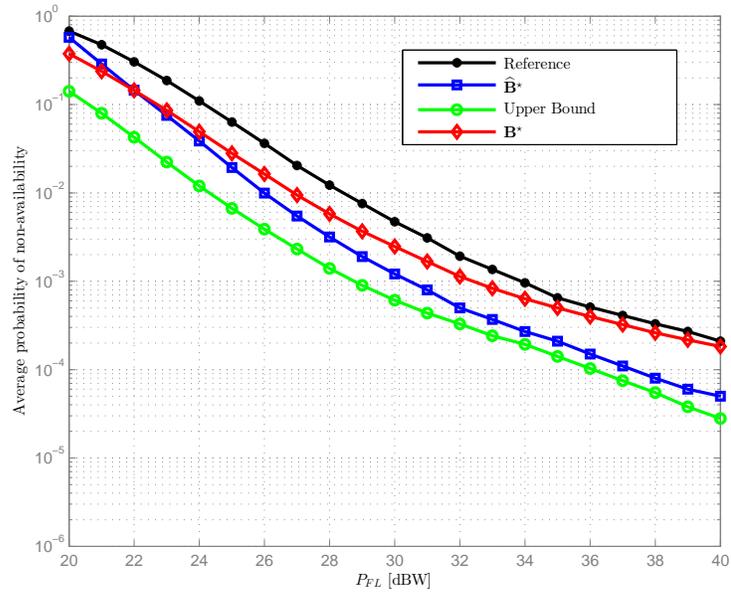}
\caption{Forward link availability.}
\label{fig4}
\end{figure}

The  results are presented for the total bandwidth  and as a function of the total available power  denoted by $P_{FL}$. Figure \ref{fig3} depicts the achieved results of spectral efficiency and Figure \ref{fig4} shows the availability of the users in the forward link. {\color{black} Clearly, the proposed techniques perform better than the benchmark system and again the robust design based on the eigenvector perturbations behaves better than the one that only considers the eigenvalues.}

\begin{figure}[tb!]
\centering
\includegraphics[scale=0.6]{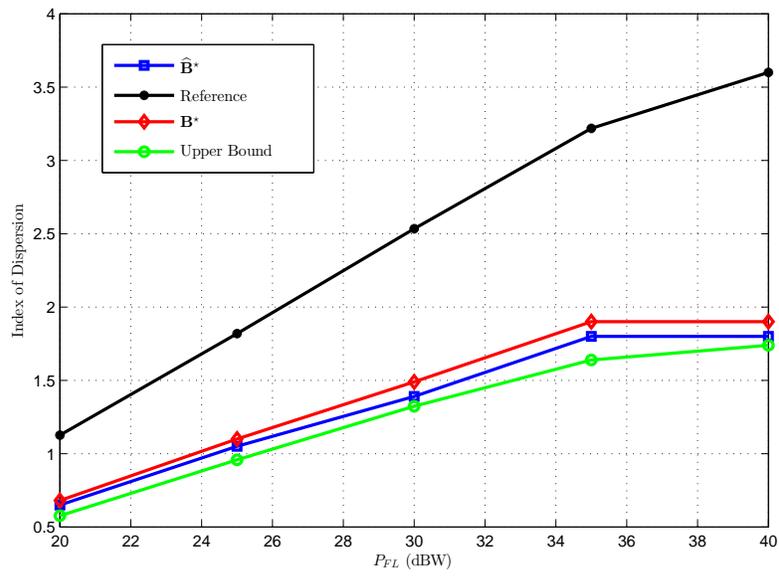}
\caption{Forward link throughput index of dispersion.}
\label{fairness_FL}
\end{figure}

\begin{figure}[tb!]
\centering
\includegraphics[scale=0.6]{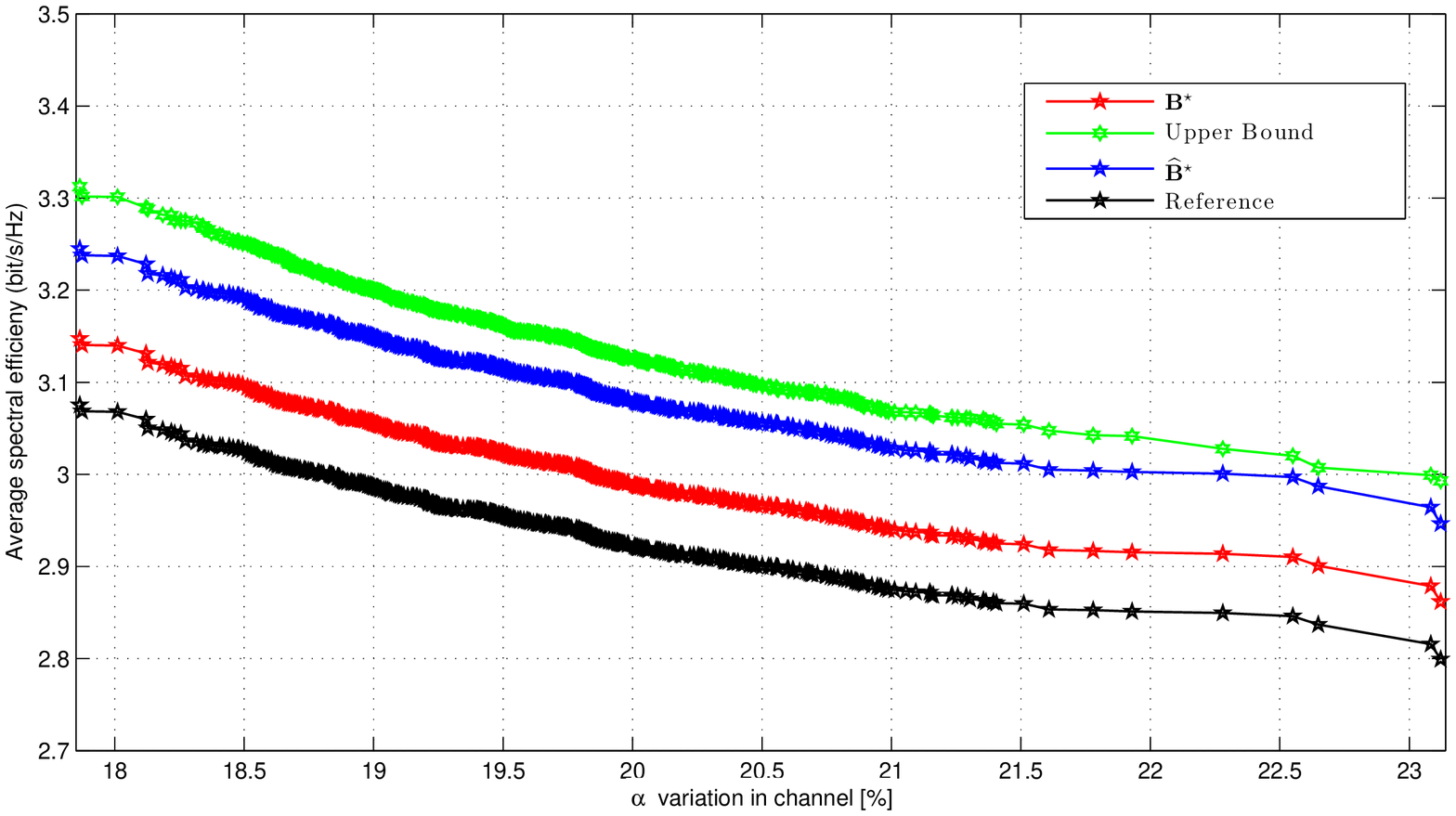}
\caption{Forward link throughput with respect to channel variations.}
\label{alpha_FL}
\end{figure}

The expected result of throughputs in Figure \ref{fig3} is justified by the availability in Figure \ref{fig4}.{\color{black} In other words,  the system with new proposed design of $\widehat{\textbf{B}}^{*}$ is closer to upper bound scenario than the reference}. Moreover, the impact of channel variations can be observed in Figure \ref{alpha_FL}.  It is clear that our proposal results in higher throughputs even when the channel variations are high. Remarkably, for the forward link the performance difference is higher than the one obtained in the return link. Note that, similar to the return link, the values of $\alpha$ are selected so that the feasibility of MSE$_{\text{FL}}$  in \eqref{222} is hold.

Finally,  the dispersion index  among users is analysed and represented in Figure \ref{fairness_FL}. For this case, the dispersion values are even higher for the reference scenario and our approach leads to higher fairness between beams.

\section{Conclusion}
This paper proposes a design of non-channel adaptive beam generation process that increases the  system throughput  compared to the conventional existing techniques in both forward and return link of a  multibeam satellite system. The design is based on an upper bound approximation of the worst case SMSE, which results to be the same for both forward and return links, leading to a large reduction of the payload complexity. The robust approximation relays on  a first perturbation model which results tighter than current robust designs. Moreover, the simulation results also have shown  the potential advantage of the considered design in order to increase  the total system throughput. As a consequence, this new approach could become a breakthrough in the design of the next satellite systems, which so far have designed the on-board beamforming only based on geographical information.


%

\appendices{

\section{}
The goal is to prove, the proposed  optimal design of $\mathbf{B}$  in \eqref{design B1} can  minimize the upper-bound of SMSE$_\text{{RL}}$ in \eqref{2}.
First, by employing  the eigenvalue decomposition of  $\breve{\mathbf{Z}}$ in \eqref{222},  problem \eqref{2} can be  rewritten as
\begin{equation}\label{478}
\min_{\mathbf{M}_{RL}}~~~~~~~\text{trace}(\mathbf{I}_K+\mathbf{ M}_{RL}\mathbf{D}_{RL}\mathbf{M}^{H}_{RL})^{-1}
\end{equation}
$~~~~~~~~~~~~~~~~~~~~~~~~~~~~~~~~~~~s.t.~~~~~~~~~~~\mathbf{M}_{RL}\mathbf{M}_{RL}^H=\mathbf{I}_K,~~~~~~~~~~~~~~~~~~~~$\\
with the following definitions
\begin{equation}\label{899}
\mathbf{M}_{RL}\triangleq\mathbf{B}\bar{\mathbf{U}},
\end{equation}
and,
\begin{math}\end{math}
 \[\mathbf{D}_{RL}\triangleq (\boldsymbol{\bar {\Sigma}}-\epsilon_{H} {\mathbf{I}_N})^{+}=\begin{pmatrix}
 \small({\boldsymbol{\bar{\Sigma}}}_{1:K }-\epsilon_H {\mathbf{I}_K})^{+} ~~~\mathbf{0}_{ K\times(N-K)}\\
  \small  ~\mathbf{0}_{(N-K)\times K} ~~~\mathbf{0}_{(N-K)\times (N-K)}
        \end{pmatrix},
\]\begin{equation}\label{652}\end{equation}
\normalsize
where $\bar{\boldsymbol{\Sigma}}$ has only $K$ non-zero eigenvalues, as $\bar{\mathbf{H}}\bar{\mathbf{H}}^H$ has rank  equal  to  $ K $. Actually, the problem   \eqref{478} can be written as
\begin{equation}\min_{\mathbf{M}_{RL}}~~~~~~\sum_{i=1}^{K} \frac{1}{1+\lambda_{i} \big(\mathbf{M}_{RL}\mathbf{D}_{RL}\mathbf{M}^{H}_{RL}\big)}
\label{479}
\end{equation}
$~~~~~~~~~~~~~~~~~~~~~~~~~~~~~~~~~~~~s.t.~~~~~~~~~\mathbf{M}_{RL}\mathbf{M}^{H}_{RL}=\mathbf{I}_K,~~~~~~~~~~~~~~~$\\
where $ {\lambda}_{i}(.) $  denotes the $i$-th largest eigenvalue of the respective matrix. Obviously, $ \mathbf{M}\mathbf{D}\mathbf{M}^H$ is a hermitian matrix whose eigenvalues  are always positive. Then, it follows that
\begin{equation}~~~~~~~g(\lambda_i)= \frac{1}{1+\lambda_{i} \big(\mathbf{M}_{RL}\mathbf{D}_{RL}\mathbf{M}^{H}_{RL}\big)}   ~~~~~~~~~            i=1, ..., K;
\label{4899}
\end{equation}
is convex function on $\lambda_{i} (\mathbf{M}_{RL}\mathbf{D}_{RL}\mathbf{M}^{H}_{RL})$. By using the theorem 3.C.1 in \cite{ ineq}, we have that
\small
\begin{equation}\phi(\boldsymbol{\lambda})=\sum_{i=1}^{K}\frac{1}{1+\lambda_{i} \big(\mathbf{M}_{RL}\mathbf{D}_{RL}\mathbf{M}^{H}_{RL}\big)}=\sum_{i=1}^{K}g\big(\lambda_{i}(\mathbf{M}_{RL}\mathbf{D}_{RL}\mathbf{M}^{H}_{RL})\big),
\label{498}
\end{equation}
\normalsize
where $ \boldsymbol{\lambda} =\big(\lambda_{1}(\mathbf{M}_{RL}\mathbf{D}_{RL}\mathbf{M}^{H}_{RL}),... ,\lambda_{K}(\mathbf{M}_{RL}\mathbf{D}_{RL}\mathbf{M}^{H}_{RL})\big)^T  $, and $\phi(.)  $  is a schur-convex function operator.
On other hand,  the theorem B.1 in \cite{ineq} proved that
\begin{equation} \mathbf{d}\prec\boldsymbol{\lambda},
\label{852}
\end{equation}
where $ \mathbf{d}(.) $ represents  $ K\times1 $ vector formed by  the diagonal elements of the matrix $ \mathbf{M}_{RL}\mathbf{D}_{RL}\mathbf{M}^{H}_{RL}$, i.e. $\mathbf{d}=   \big(d_{1}(\mathbf{M}_{RL}\mathbf{D}_{RL}\mathbf{M}^{H}_{RL}),... ,d_{K}(\mathbf{M}_{RL}\mathbf{D}_{RL}\mathbf{M}^{H}_{RL})\big)^{T} $. Finally,  combining of \eqref{852} with the schur convexity of $\phi(.)$, we have that $\phi(\mathbf{d})\leq\phi(\boldsymbol{\lambda})$, i.e.
\begin{equation}\sum_{i=1}^{K}\frac{1}{1+d_{i} \big(\mathbf{M}_{RL}\mathbf{D}_{RL}\mathbf{M}^{H}_{RL}\big)}\geq\sum_{i=1}^{K}\frac{1}{1+\lambda_{i} \big(\mathbf{M}_{RL}\mathbf{D}_{RL}\mathbf{M}^{H}_{RL}\big)}.
\label{855}
\end{equation}
Moreover, the equality in \eqref{855} is reached whenever $\mathbf{M}_{RL}\mathbf{D}_{RL}\mathbf{M}^{H}_{RL}$ is diagonal. To this end, it is clear that $\mathbf{M}$ has to be diagonal such that
\begin{equation} \mathbf{M}_{RL}=[\mathbf{I}_K  ~~~~~ \mathbf{0}_{K\times (N-K)}].
\label{856}
\end{equation}
Given \eqref{899}, it  implies that $\mathbf{B}$
has to be made of the  $ K$ first rows of the matrix  $\bar{\mathbf{U}}^{H}$, that is
\begin{equation} \mathbf{B}=\bar{\mathbf{L}}_{1:K}^H,
\label{857}
\end{equation}
and concludes the proof.

{\color{black}
\section{}

The starting point of the derivation is the upper bound obtained when only considering the eigenvalues variation
\begin{equation}
\bar{\mathbf{U}}_s\left(\bar{\boldsymbol{\Sigma}} - \epsilon_H\mathbf{I} \right) \bar{\mathbf{U}}^H_s,
\end{equation}
where for this case we additionally consider the perturbation on the eigenvectors as
\begin{equation}
\left( \bar{\mathbf{U}}_s + \Delta\mathbf{U}_s\right)\left(\bar{\boldsymbol{\Sigma}} - \epsilon_H\mathbf{I} \right) \left(\bar{\mathbf{U}}_s + \Delta\mathbf{U}_s\right)^H.
\end{equation}
In \cite{liu8} it is presented that the perturbation on the eigenvectors take the form of
\begin{equation}
\Delta\mathbf{U}_s = \bar{\mathbf{U}_s}\mathbf{R} + \bar{\mathbf{U}_n}\bar{\mathbf{U}_n}^H\Delta\mathbf{Z}\bar{\mathbf{U}_s}\bar{\boldsymbol{\Sigma}}_s^{-1}
\end{equation}
where
\begin{equation}
\mathbf{R} = \mathbf{D}\cdot \left(\mathbf{U}_s^H\Delta\mathbf{Z}\mathbf{U}_s\bar{\boldsymbol{\Sigma}} + \bar{\Sigma}\mathbf{U}_s^H\Delta\mathbf{Z}^H\mathbf{U}_s \right),
\end{equation}
and the $g,f$-th entry of $\mathbf{D}$ is
\begin{equation}
\lambda_f - \lambda_g,
\end{equation}
for $f\neq g$ and $\lambda_f$ for $f=1,\ldots,N$ denote the eigenvalues of $\bar{\mathbf{H}}\bar{\mathbf{H}}^H$.
Considering that 
\begin{equation}
\Delta\mathbf{Z} \leq \epsilon_H\mathbf{I},
\end{equation}
the following inequality holds
\begin{equation}
\Delta\mathbf{U}_s \leq \bar{\mathbf{U}}_s\mathbf{R} + \epsilon_H\bar{\mathbf{U}}_n\bar{\mathbf{U}}_n^H\bar{\mathbf{U}}_s\bar{\boldsymbol{\Sigma}}_s^{-1}.
\end{equation}
Additionally, we have that
\begin{equation}
\mathbf{U}_s^H\Delta\mathbf{Z}\mathbf{U}_s\bar{\boldsymbol{\Sigma}} + \bar{\Sigma}\mathbf{U}_s^H\Delta\mathbf{Z}^H\mathbf{U}_s \leq \epsilon_H\mathbf{U}_s^H\mathbf{U}_s\bar{\boldsymbol{\Sigma}} + \epsilon_H\bar{\boldsymbol{\Sigma}}\mathbf{U}_s^H\mathbf{U}_s.
\end{equation}
The following lemma is required for obtaining the result

\textbf{Lemma 2} \emph{For any semidefinite positive matrices $\mathbf{A}$,$\mathbf{K}$,$\mathbf{C}$, and $\mathbf{K}\leq \mathbf{C}$, it holds that}
\begin{equation}
\mathbf{A} \cdot \mathbf{K} \leq \mathbf{A} \cdot \mathbf{C}.
\end{equation}
\begin{proof}
See Theorem 17 of \cite{ham78}.
\end{proof}

With this last result it is possible to write the following
\begin{equation}
\Delta\mathbf{U}_s \leq \bar{\mathbf{U}}_s\widehat{\mathbf{R}} + \epsilon_H\bar{\mathbf{U}}_n\bar{\mathbf{U}_n}^H\bar{\mathbf{U}}_s\bar{\boldsymbol{\Sigma}}_s^{-1},
\end{equation}
where
\begin{equation}
\widehat{\mathbf{R}} = \mathbf{D}\cdot \left(\epsilon_H\mathbf{U}_s^H\mathbf{U}_s\bar{\boldsymbol{\Sigma}} + \epsilon_H\bar{\Sigma}\mathbf{U}_s^H\mathbf{U}_s \right).
\end{equation}
}
}

{\color{black}
\section*{Acknowledgement}

The authors would like to thank the anonymous reviewers whose comments extremely increase the quality of the paper.
}
\ifCLASSOPTIONcaptionsoff
  \newpage
\fi



\bibliographystyle{IEEEtran}
\bibliography{IEEEabrv}
%

%

%
%
%




\end{document}